\documentclass[twocolumn,showpacs,%
  nofootinbib,aps,superscriptaddress,%
prd,notitlepage,showkeys,groupedaddress,10pt]{revtex4-1}

\usepackage{amssymb}
\usepackage{amsmath}
\usepackage{graphicx}
\usepackage{dcolumn}
\usepackage{subcaption}
\include{float}
\usepackage{hyperref}

\begin{document}

\title{Quench Dynamics in Confined 1+1-Dimensional Systems}
\author{Dalit Engelhardt}
\email{engelhardt@physics.ucla.edu}
\affiliation{\vspace{0.2in}Department of Physics and Astronomy, University of California, Los Angeles, CA 90095, USA\vspace{0.1in}\\Institute for Theoretical Physics, University of Amsterdam, Science Park 904, Postbus 94485, 1090 GL Amsterdam, The Netherlands}

\begin{abstract}
We present a framework for investigating the response of conformally-invariant confined 1+1-dimensional systems to a quantum quench. While conformal invariance is generally destroyed in a global quantum quench, systems that can be described as or mapped to integrable deformations of a CFT may present special instances where a conformal field theory-based analysis could provide useful insight into the non-equilibrium dynamics. We investigate this possibility by considering a quench analogous to that of the Quantum Newton's Cradle experiment~\cite{Kinoshita} and demonstrating qualitative agreement between observables derived in the CFT framework and those of the experimental system. We propose that this agreement may be a feature of the proximity of the experimental system to an integrable deformation of a $c=1$ CFT.
\end{abstract}

\maketitle

The analytical modeling of the out-of-equilibrium behavior of systems
subjected to a quantum quench --- a sudden change in the system's Hamiltonian
parameters --- remains a challenging problem in all but a limited number
of simple cases. Of particular interest are systems that do not exhibit
simple relaxation to a thermal state following a quantum quench; while
generically we expect systems subjected to sudden changes to eventually
thermalize and reach an equilibrium state as a result of interactions,
the past decade has seen accummulating experimental evidence for and
theoretical studies on one-dimensional systems that do not equilibrate
to a simple thermal state, but that instead appear to retain memory
of their initial state (for reviews see~\cite{PolkovnikovReview,DaleyReview,EisertReview}).
Integrability has been shown to inhibit thermalization~\cite{Rigol},
and such behavior in experimental systems is often attributed to the
proximity of these systems to an integrable point.

The connections between systems that are integrable and those that
are conformally-invariant have been subject
to ongoing investigation. In particular, certain integrable field
theories can be obtained from massive deformations of particular models
of conformal field theory (CFT)~\cite{Zam}, rendering the understanding
of thermalization within a CFT framework a potentially powerful tool
for testing some of the ideas arising in the study of the connections between conformal invariance and integrability~\cite{IntegrableStructure}.
In principle, if it is known how a particular integrable model arises
as a perturbation at a conformal fixed point, conformal perturbation
theory (see, e.g.,~\cite{ZamPerturbedCFT}) can be used to compute observables of the integrable theory
up to arbitrary order. While this approach may often become computationally cumbersome beyond the lowest orders, it raises the question of whether out-of-equilibrium
analyses of certain CFTs may shed light on the post-quench behavior
of related integrable models.

In this Letter we address this question by providing an example of a realistic near-integrable quenched system that we show exhibits the behavior characteristic of a conformally-invariant system. The system that we consider is that of the ``Quantum Newton's Cradle'' experiment~\cite{Kinoshita}, in which an effectively one-dimensional system of interacting harmonically-confined bosons was split into two oppositely-moving momentum groups; following this quench, the system failed to demonstrate any apparent thermalization within experimental time scales. While some experimental effects, such as the presence of a confining trapping potential in the setup, may introduce weak integrability-breaking effects, the system is believed to be well-described by the integrable Lieb-Liniger model~\cite{LiebLiniger} of delta-interacting bosons, and the failure of the experimental system to thermalize has been attributed to the integrability of this model.

To motivate the relation to the CFT picture, we note that the non-relativistic Lieb-Liniger model can be exactly mapped to the relativistic sinh-Gordon model in an appropriate limit. In particular, under this mapping the S-matrix and Lagrangian of the two models coincide~\cite{Kormos1,Kormos2}. The sinh-Gordon model is a massive integrable deformation of a free scalar field Lagrangian, and correlation functions in this model can in principle be computed order-by-order in a conformal perturbation expansion. 

We proceed as if this system were a $c=1$ CFT, which is an accurate effective description of the Lieb-Liniger model in the limit of either low momenta or hard-core boson interactions (the latter which map to free fermions~\cite{Girardeau}). This amounts to neglecting higher order terms in a perturbative expansion of correlation functions of the sinh-Gordon model and hence its non-relativistic Lieb-Liniger limit. An important issue when truncating such a perturbative expansion is whether higher-order perturbative effects, which may not qualitatively change the behavior in equilibrium, could have significant effect in a non-equilibrium setting on the asymptotic (long-time) behavior of observables. As we show, this does not appear to be the case in a qualitative analysis; we comment on this and suggest how a quantitative analysis may be performed in order to detect potential deviations.

In the free boson CFT analyzed here, the harmonic confinement of the system implies that (up to an overall rescaling) there is a full equivalence up to a phase lag of half the system's size between the position-space energy density expectation value (given by $\langle T_{tt}\rangle$) and the momentum-space expectation value. The latter is the CFT observable corresponding to the momentum distributions observed in the experiment. Although the experimental setup was in principle not limited to the low momenta or hard-core interactions regimes, we show that the experimental momentum distributions and this CFT observable qualitatively agree.

Methods for analyzing quenches in a CFT via a boundary state approach were proposed in ~\cite{CardyCalabrese1,CardyCalabrese2,Cardy2014}.
In these constructions\footnote{See also e.g.~\cite{Gambassi1, Gambassi2,Tonni} for recent work building on these methods.} the system, with Hamiltonian $H$, is prepared at $t=0$ in an initial state $\left|\psi_{0}\right\rangle $, which is an eigenstate of a different Hamiltonian $H_{0}$. In an analytically-continued Euclidean version
of the theory this initial state can be interpreted as a Euclidean ``boundary state'' $\left|B\right\rangle $ that encodes the initial conditions
of $\left|\psi_{0}\right\rangle $. The state of the system at any later time is then given by $\left|\psi(t)\right\rangle =e^{-iHt}\left|\psi_{0}\right\rangle $,
and correlation functions of observables can be computed in the
boundary state $\left|B\right\rangle $, followed by an analytic continuation to Lorentzian time. Since conformal boundary states
are non-normalizable, the actual boundary state $\left|B\right\rangle $
is taken to be at a certain RG distance --- the ``extrapolation length''
$\tau_{0}$ --- from the conformal boundary state $ $$\left|B\right\rangle _{CFT}$. 

The introduction of this extrapolation length gives the boundary state
a width of $2\tau_{0}$: rather than evolving from a single Euclidean
boundary state, the time-evolution is from the center $\tau=0$ of
a slab, or strip, whose top and bottom boundaries at $\tau=\pm\tau_{0}$
correspond to the same initial state. Real-time $t$ correlation functions
are then obtained by evolving observables from $\tau=0$ and analytically
continuing $\tau\rightarrow-it$. Correlation functions in this setup
in a 2D boundary CFT (BCFT)~\cite{CardyBCFT} can then be computed by making use of the conformal transformation
that maps the strip to the half-plane.

\begin{figure}
  \centering
    \includegraphics[width=1\linewidth]{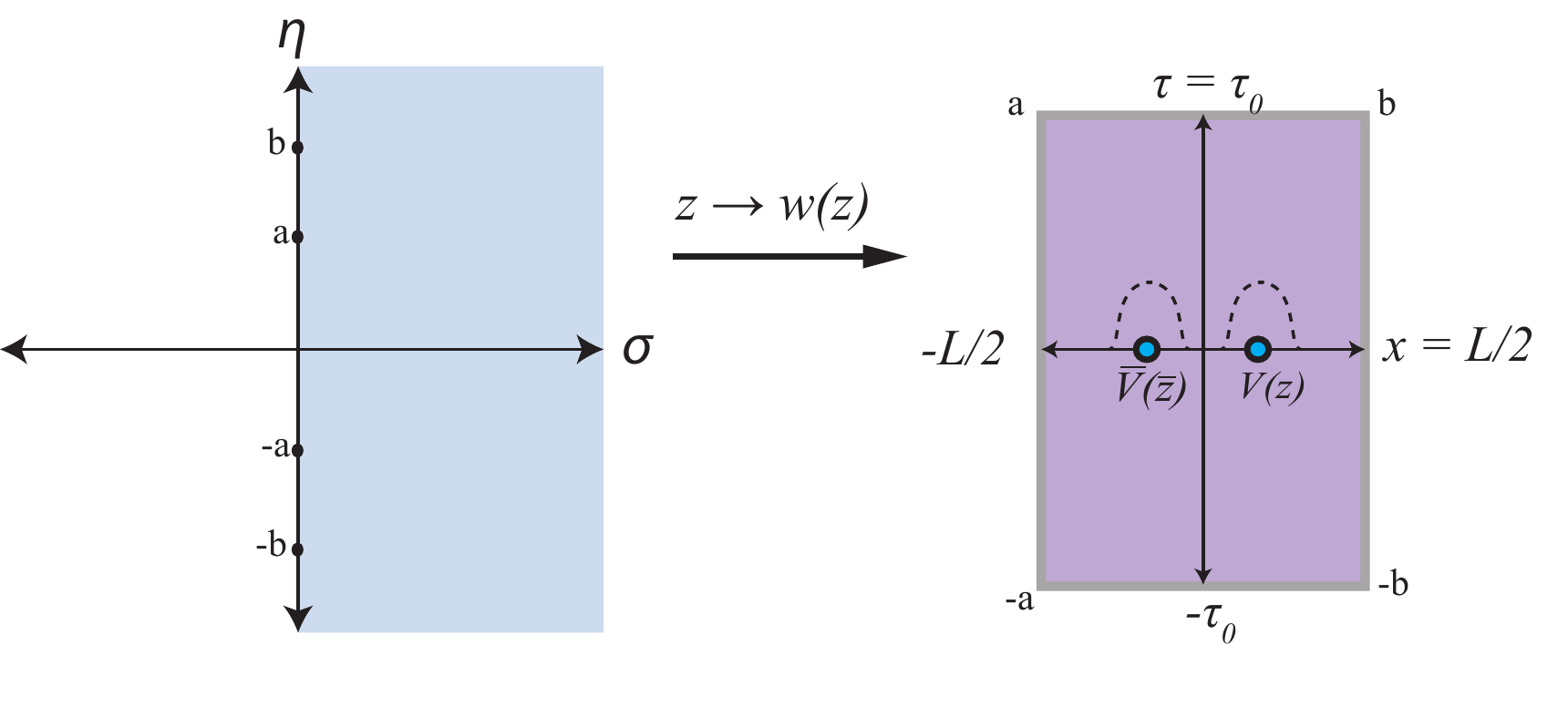}
\caption{\small{The Schwarz-Christoffel tranformation taking the right-half plane to the rectangle with chiral and antichiral vertex operators inserted on the vertical midline, $\tau=0$}}
\label{fig:rectangle}
\end{figure}

\vspace{0.1in}
\textit{Spatial confinement ---} To make contact with the realistic system we modify this formalism
to account for systems on a finite interval. We introduce the spatial confinement by adding boundaries
along the spatial direction such that the length of the system is
now given by $L$. The resulting boundary state geometry is therefore
that of a rectangle of length $L$ and height $2\tau_{0}$ (Fig.~\ref{fig:rectangle})
which also conformally maps to the half-plane. The transformation
to the right-half plane~\cite{KunsMarolf} is given by an elliptic
Jacobi function\footnote{The modification by an additive constant here from~\cite{KunsMarolf} centers the resulting rectangle on the origin of the transformed coordinates.}
\begin{equation}
w\rightarrow z(w)=\left.\text{sc}\left(\frac{K_{1}\left(k\right)}{L}\left(w+\frac{L}{2}\right),k\right)\right].\label{eq:jacobielliptic}
\end{equation}
where $k$ is the elliptic modulus, $k\in\left[0,1\right]$ and $K_{1}(...)$
is the complete elliptic integral of the first kind. Its inverse is
a Schwarz-Christoffel transformation~\cite{Driscoll} given by the
elliptic integral of the first kind 
\[
w(z)=\frac{L}{K_{1}(k)}F\left(\tan^{-1}(z),k\right)-\frac{L}{2}
\]
and that maps a set of designated points $y=\pm 1,\pm 1/\sqrt{1-k}$
on the imaginary ($x=0$) axis to the vertices of a
rectangle as shown in Fig.~\ref{fig:rectangle} with height
$2\tau_{0}=2K_{1}\left(\sqrt{1-k}\right)/K_{1}\left(k\right)$ where $\tau_0$ is the
extrapolation length of the previous section. The limit of $k\rightarrow0$ corresponds to the infinite-height rectangle
(strip) and $k\rightarrow1$ is the limit of zero height. The mapping
(\ref{eq:jacobielliptic}) is doubly-periodic (i.e. with one period equal to $2L$)
in the (real) argument\footnote{As a result of the continuation to Lorentzian time all arguments considered here are real.} for $\tau_0>0$;
this is a feature of the open reflective boundary conditions that
it imposes, and as a result observables in this geometry will
display periodic returns to their initial values, though the period may change with the number of insertions. This makes this choice of boundary state geometry particularly well-suited for modeling harmonically-confined systems. The special case
of the ground-state expectation value, the Casimir energy, has periodicity
$L$ as the Jacobi elliptic functions only appear squared~\cite{KunsMarolf}.

\vspace{0.1in}
\textit{Vertex operator insertions ---} We introduce excitations in this setup by restricting to $c=1$ CFT and considering local vertex operator insertions. We note that the formalism described here carries over with minor modifications to $c<1$ minimal models through the addition of screening charges to all correlation functions. We thus consider the free boson action
\begin{equation}
S=\frac{g}{2}\int d^{2}x\left(\nabla\phi(\vec{x})\right)^{2},\label{eq:gaussianAction}
\end{equation}
where $\phi$ is a bosonic field, which we take to be compactified on a circle of radius
$R$, $\phi\sim\phi+2\pi R$, which we henceforth set to $R=1$. Highest-weight states are given by the action of vertex
operators on the vacuum state, $\left|n,m\right\rangle =\lim_{z,\bar{z}\rightarrow0}V_{nm}(z,\bar{z})\left|0\right\rangle$, where $V_{nm}(z,\bar{z})=V_{nm}(z)\otimes V_{\bar{n}\bar{m}}(\bar{z})$, and the chiral and antichiral vertex operators are given respectively as $V_{nm}(z)=:e^{i\alpha_{nm}\phi(z)}:$
and $V_{\bar{n}\bar{m}}(\bar{z})=:e^{i\bar{\alpha}_{nm}\bar{\phi}(\bar{z})}:$,
where $m$ is the winding number and $n$ is the wave number~\cite{diFrancesco}.

Their holomorphic and antiholomorphic conformal dimensions are given by $h_{nm}=\frac{\alpha_{nm}^{2}}{8\pi g}=\frac{1}{8\pi g}\left(n+\frac{m}{2}\right)^{2}$
and $\bar{h}_{nm}=\frac{\bar{\alpha}_{nm}^{2}}{8\pi g}=\frac{1}{8\pi g}\left(n-\frac{m}{2}\right)^{2}.$
A Luttinger liquid CFT, for instance, is obtained
by setting the normalization $g=K$ in (\ref{eq:gaussianAction}),
where $K$ is the Luttinger parameter. The bosonic field $\phi$ now
represents propagating density fluctuations and
is related to the dual variable $\theta$ under the T-duality transformation
$\phi\leftrightarrow\theta$ and $K\leftrightarrow1/K,\: n\leftrightarrow m$.
For the comparison with the experimental data in the following section
we will set $g=K$ in subsequent calculations.

\vspace{0.1in}
\textit{Split-momentum quench ---} We implement the split momentum quench as a boundary state given by a pair
of chiral $V_{nm}(z)$ and antichiral $V_{\bar{n}\bar{m}}(\bar{z})$
vertex operators of the compactified free boson (Fig.~\ref{fig:rectangle})  together with Dirichlet boundary conditions in conjunction with the elliptic Jacobi mapping. The opposite-chirality vertex operators act to excite the ground state in analogy to the experimental setup of two excited oppositely-moving clouds of bosons. In the CFT analogy, each such cloud is represented as a peak given by the location of the vertex operator; in reality the clouds have a certain spread, and we later discuss how this spread can be accounted for in the CFT analysis. The Dirichlet conditions in conjunction with the inherent periodicity of the conformal mapping are implemented to mimick the harmonic trapping potential of the experimental setup.

The Dirichlet condition selects the type of boundary states allowed~\cite{Ishibashi1,Ishibashi2,Cardy89}, which are given by~\cite{HsuFradkin,Oshikawa}
\begin{equation}
\left|\left|D\right\rangle \right\rangle =\left(4\pi K\right)^{-\frac{1}{4}}\sum_{n\in\mathbb{Z}}e^{i\frac{n\phi_{0}}{\sqrt{4\pi K}}}\left|\left.(n,0)\right\rangle \right\rangle _{D}\label{eq:DirichletBoundaryState}
\end{equation}
where $\phi_{0}$ is canonically conjugate to the zero mode of the
free boson and takes values in a circle of radius $R=1$. The normalization
$\left(4\pi K\right)^{-\frac{1}{4}}$ ~\cite{Venuti} is the g-factor (boundary entropy)~\cite{AffleckLudwig} for the Dirichlet
boundary condition for the action (\ref{eq:gaussianAction}) with
$g=K$. Unlike in the boundary-less case, expectation values of primary
operators do not in general vanish in a BCFT; in the case of the compactified
boson the expectation value can be obtained from the boundary states
above as
\begin{equation}
\left\langle V_{nm}(z,\bar{z})\right\rangle _{D}=\frac{1}{\sqrt{K_{0}}}e^{i\frac{n\phi_{0}}{K_{0}}}\left|z-\bar{z}\right|^{-\frac{n^{2}}{K_{0}^{2}}}\label{eq:onepoint}
\end{equation}
where $K_{0}=\sqrt{4\pi K}$.

The energy density expectation value $\left\langle \psi_{0}\left|T_{tt}\left(t,x\right)\right|\psi_{0}\right\rangle $
at time $t$ for the initial state of the split-momentum quench is
given, upon analytic continuation $t\rightarrow i\tau$, by 
\begin{equation} 
\frac{1}{2\pi}\left\langle \left\langle D_{\text{rec}}\left|\left|\left(T(w)+\bar{T}(\bar{w})\right)V_{nm}\left(w'\right)V_{\bar{n}\bar{m}}\left(\bar{w}'\right)\right|\right|D_{\text{rec}}\right\rangle \right\rangle 
\label{eq:Tttexplained}
\end{equation}
where $\left|\left|D_{\text{rec}}\right\rangle \right\rangle $ is
the boundary state state (\ref{eq:DirichletBoundaryState})
following the conformal transformation to the rectangle, and we have used the decomposition of the energy
density as the sum of holomorphic and antiholomorphic components. Coordinates
on the rectangle will be denoted by $w=x+i\tau$ and on the half-plane
by $z=\sigma+i\eta$. Recall that it is the Euclidean time coordinate of the
stress tensor, $T_{\tau\tau}$, rather than the time coordinates of
the vertex operators, that is analytically continued to Lorentzian
time. The coordinates $w'=x'+i\tau'$, where $\tau'=0$, denote the
location of the vertex operator insertion on the rectangle. The equivalence
of (\ref{eq:Tttexplained}) with the time-evolved
expectation value of the energy density from the given initial state
can be understood by noting that the right-hand side can be formally
expressed as a Euclidean path integral with an operator insertion.

The expectation value~(\ref{eq:Tttexplained}) can be computed by
conformally transforming both the vertex operators and the stress
tensor to the half-plane. The vertex operators transform under the conformal transformation
as primary fields, $V_{nm}\left(w'\right)=\left(\frac{dw'}{dz'}\right)^{-h_{nm}}V_{nm}\left(z'\right)$
and $V_{nm}\left(\bar{w}'\right)=\left(\frac{d\bar{w}'}{d\bar{z}'}\right)^{-\bar{h}_{nm}}V_{nm}\left(\bar{z}'\right)$,
whereas the stress tensor acquires the anomalous Casimir term, $T(w)=\left(\frac{dw}{dz}\right)^{-2}T(z)+\frac{c}{12}\{z;w\},$
where $\{z;w\}=\frac{\left(d^{3}z/dw^{3}\right)}{\left(dz/dw\right)}-\frac{3}{2}\left(\frac{d^{2}z/dw^{2}}{dz/dw}\right)^{2}$
is the Schwarzian derivative. In the absence of spatial boundaries,
i.e. the infinite strip limit of $k\rightarrow0$, the Schwarzian
derivative term is equal to the constant strip Casimir enery. The
Casimir term produced by the Jacobi elliptic transformation (\ref{eq:jacobielliptic})
for $k>0$ is not a constant, and it has a significant qualitative
effect on the energy distribution. Employing the Ward identity on
the upper-half plane~\cite{CardyBCFT}
\vspace{-0.1in}
\begin{widetext}
\[
\left\langle T(z)V_{nm}(z')V_{\bar{n}\bar{m}}(\bar{z}')\right\rangle \sim\left(\frac{\partial_{z'}}{z-z'}+\frac{h_{nm}}{(z-z')^{2}}+\frac{\partial_{\bar{z}'}}{z-\bar{z}'}+\frac{\bar{h}_{nm}}{(z-\bar{z}')^{2}}\right)\left\langle V_{nm}(z')V_{\bar{n}\bar{m}}(\bar{z}')\right\rangle 
\]
we arrive at the expression for (\ref{eq:Tttexplained})
\begin{equation}
\begin{aligned}\left\langle T_{\tau\tau}(w,\bar{w})\right\rangle  & =\frac{1}{2\pi\sqrt{K_{0}}}\left(\frac{dw'}{dz'}\frac{d\bar{w}'}{d\bar{z}'}\right)^{-\frac{n^{2}}{2K_{0}^{2}}}e^{i\frac{n\phi_{0}}{K_{0}}}\left|z'-\bar{z}'\right|^{-\frac{n^{2}}{K_{0}^{2}}}\\
 & \times\left\{ \frac{n^{2}}{2K_{0}^{2}}\left(\frac{dw}{dz}\right)^{-2}\left[\frac{2}{\left|z'-\bar{z}'\right|}\left(\frac{-1}{z-z'}+\frac{1}{z-\bar{z}'}\right)+\frac{1}{\left(z-z'\right)^{2}}+\frac{1}{\left(z-\bar{z}'\right)^{2}}\right]+\frac{1}{12}\left\{ z(w),w\right\} +a.h.\right\} 
\end{aligned}
\label{eq:TtautauFull}
\end{equation}

\end{widetext}
where $a.h.$ refers to the antiholomorphic part of the expression,
i.e. $z\rightarrow\bar{z}$, $w\rightarrow\bar{w}$, and as a result of the Dirichlet boundary condition we have set
$h=h_{nm}=\bar{h}_{nm}=\frac{n^{2}}{2K_{0}^{2}}$, where as before
$K_{0}^{2}=4\pi K$, and made use of (\ref{eq:onepoint}) in computing the chiral-antichiral vertex operator correlator. We stress
that the coordinates $z$ in (\ref{eq:TtautauFull}) must be read
as functions of the rectangle coordinates $w$, related via (\ref{eq:jacobielliptic}).

Since the transformation (\ref{eq:jacobielliptic}) is from the right-half
plane, the antiholomorphic coordinates $\bar{z}$ are rotated from
the usual upper-half plane ones, i.e. $\bar{z}=-z^{*}$. Finally, the Lorentzian energy expectation value $\left\langle T_{tt}(t,x)\right\rangle $
is obtained via a Wick rotation $w=x+i\tau\rightarrow x+t$ and $\bar{w}=-x+i\tau\rightarrow-x+t$.

We note that while there appear to be four divergences in (\ref{eq:TtautauFull})
for all times $t>0$, in fact two of these divergences fall outside
of the rectangle boundaries at any given time, so that there are effectively
only two remaining divergences. These divergences oscillate within the confines of the system, coinciding twice within each period of the full energy distribution $\left\langle T_{tt}(t,x)\right\rangle $, which is $2L$ as a result of (\ref{eq:jacobielliptic}). These divergences are a feature
of an analysis that -- despite the conformal transformation to a finite
geometry -- has been carried out in the thermodynamic limit. They are a consequence of the divergence of the correlation
length in the thermodynamic limit: since the system that we consider
here is finite of length $L,$ the divergences are rounded off owing
to the effects of finite size scaling~\cite{CardyFiniteSize} (see Appendix~\ref{sec:regulation} for regulation scheme).

The physical picture that emerges from (\ref{eq:TtautauFull}) is
that of the non-constant Casimir term, $\frac{c}{12}\left\{ z(w),w\right\} +a.h.$,
owing to the special type of confinement imposed, competing in strength
with the two oscilating bumps (regulated divergences) of the terms involving the momentum excitations given by the vertex operator insertions. The strength of these bumps is given
by the vertex operators' conformal dimension $h=\frac{n^{2}}{2K_{0}^{2}}$, so that the relative
strength of these momentum packets to the Casimir term increases with lower values of the Luttinger
parameter $K$. Since decreasing values of $K$ correspond to increasing values of the Lieb-Liniger parameter $\gamma$~\cite{Cazalilla04}, their strength increases with $\gamma$. 

The elliptic Jacobi transformation with Dirichlet boundary conditions thus mimicks the behavior observed in the experiment~\cite{Kinoshita}, where the two momentum packets repeatedly oscillate within the harmonic trap with a periodicity such that the two packets coincide twice per period. As a result of the harmonic symmetry of the setup, in the case of the non-interacting system assumed in the CFT analysis a direct comparison is possible between position-space and momentum-space distributions: at any time $t_p$ the momentum-space distribution is seen to be equivalent (up to an overall scale factor) to the position-space distribution at $t_x=t_p+L/2$. In the CFT analysis we have assumed that the two packets are highly localized; a realistic spread in the momentum may be accounted for by shifting the spatial coordinate away from the endpoints of the interval and
closer to the middle. 

\begin{figure*}
\centering
\begin{tabular}{cccc}
  \includegraphics[width=0.23\textwidth]{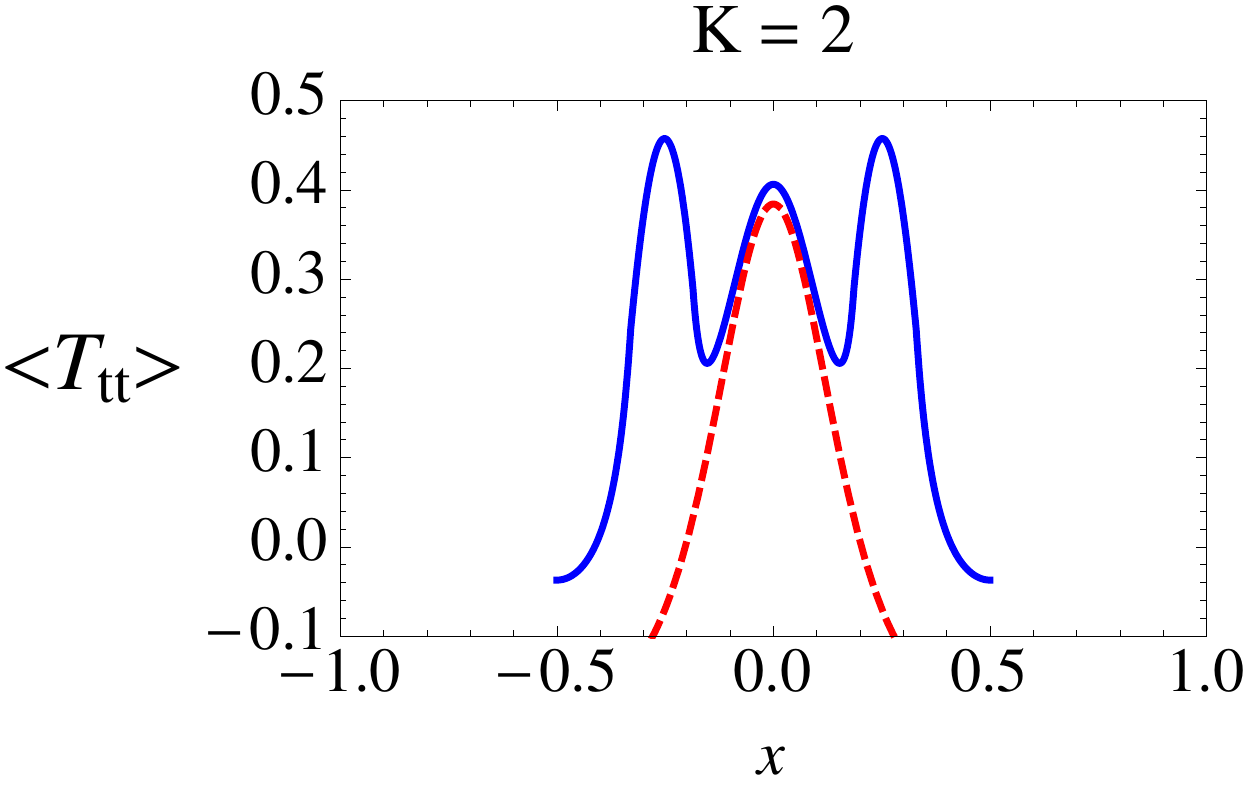}
  \includegraphics[width=0.23\textwidth]{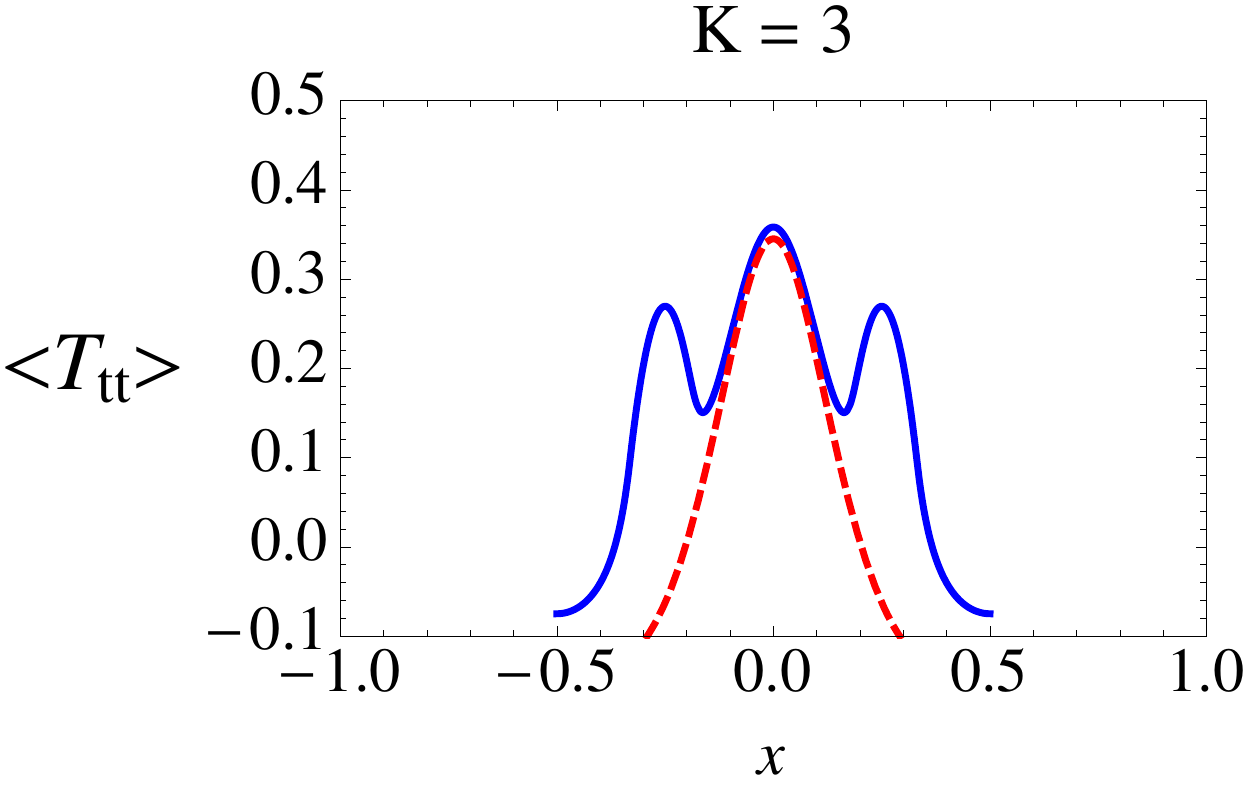}
  \includegraphics[width=0.23\textwidth]{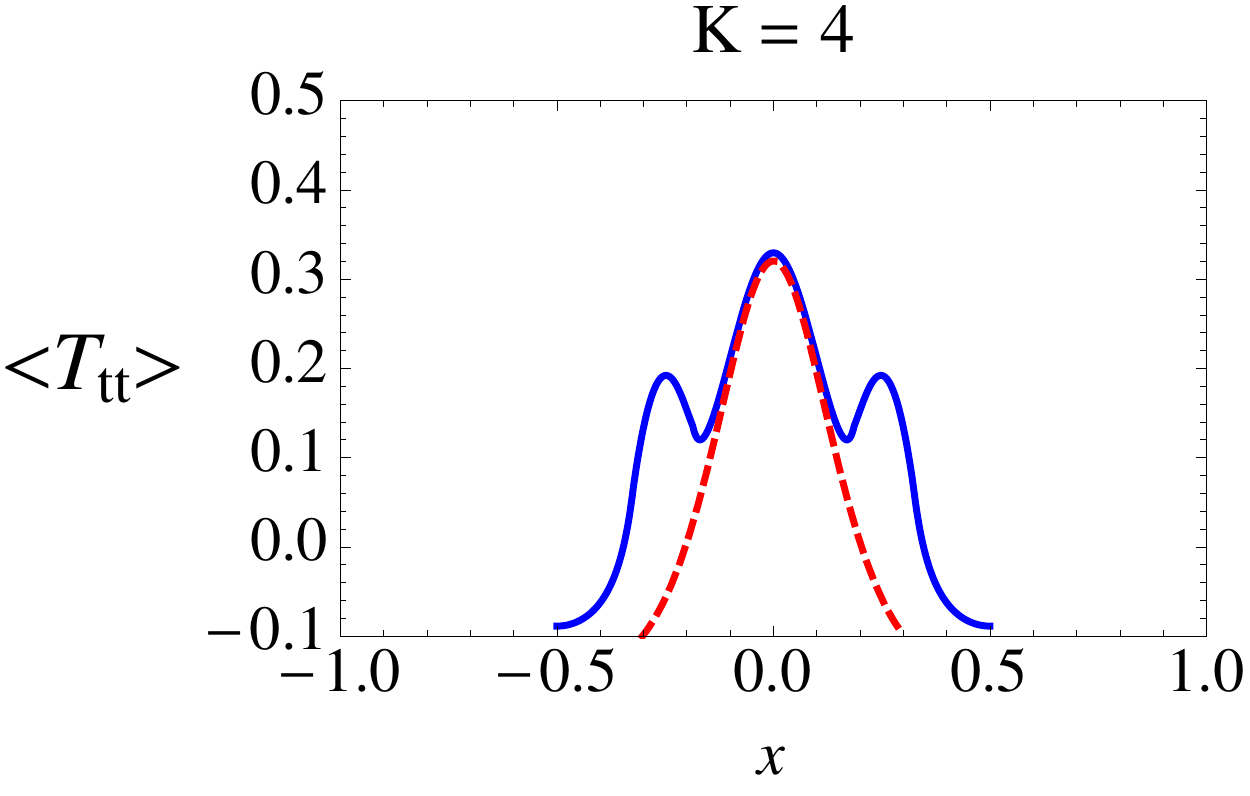}
  \includegraphics[width=0.23\textwidth]{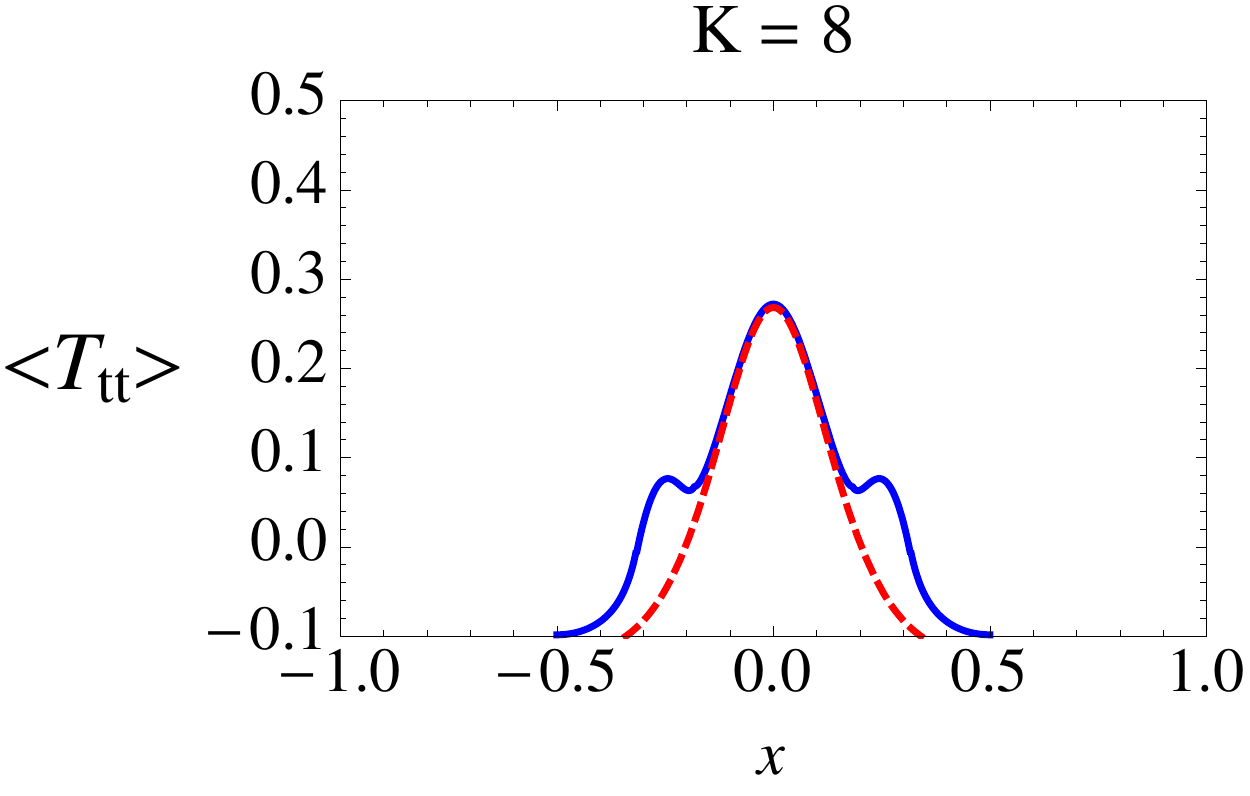}\\
 \includegraphics[width=0.23\textwidth]{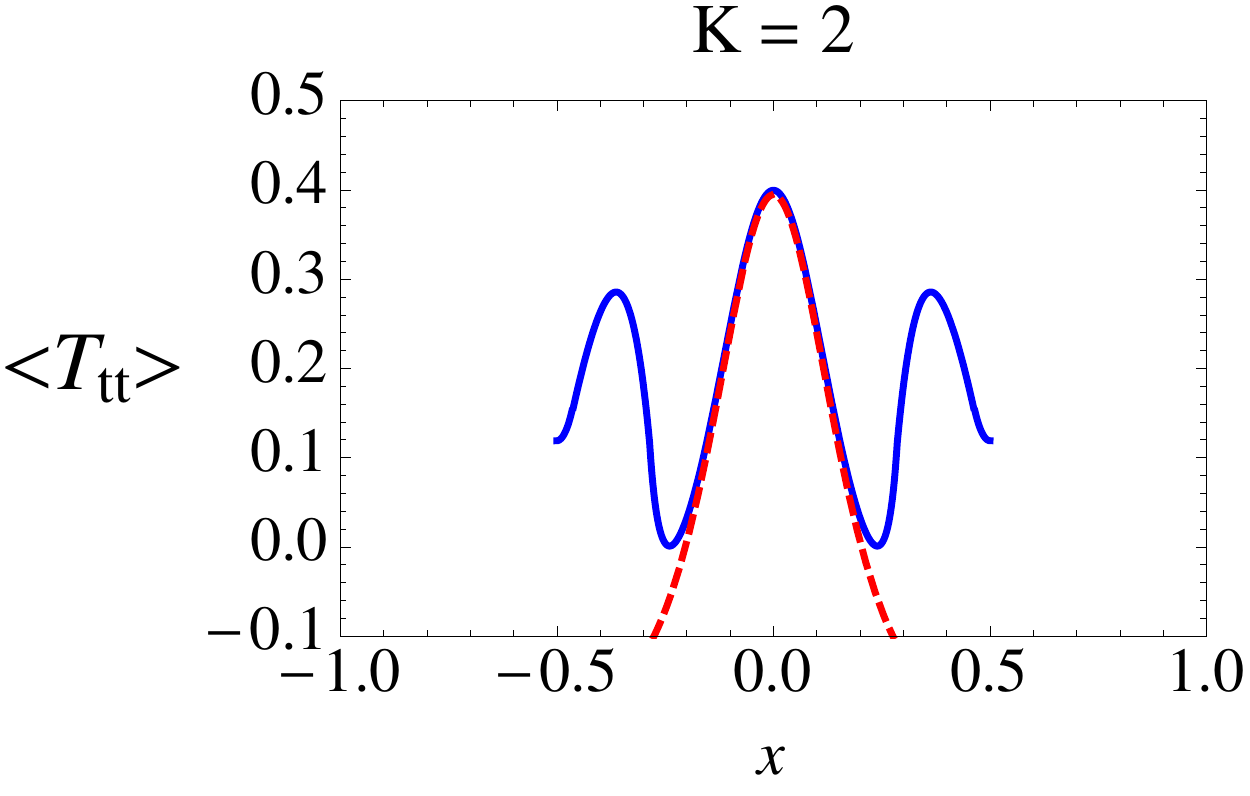}
  \includegraphics[width=0.23\textwidth]{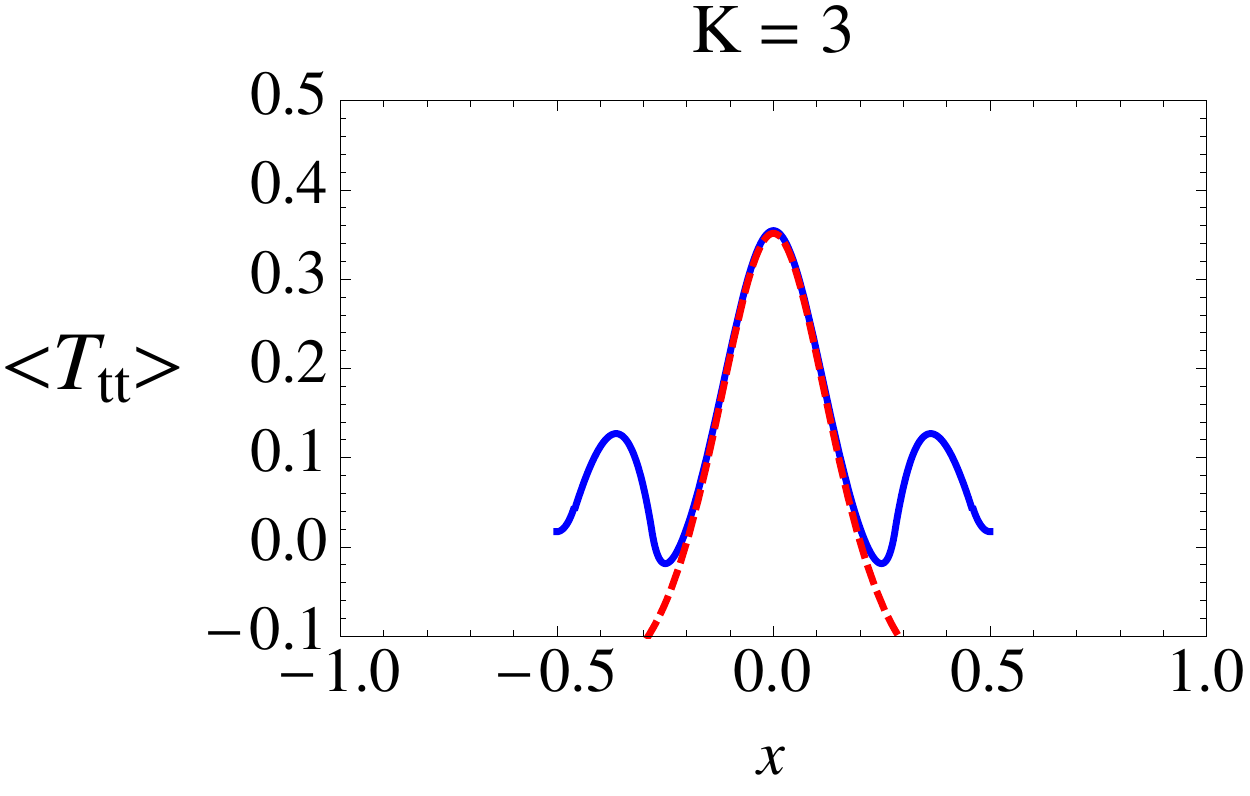}
  \includegraphics[width=0.23\textwidth]{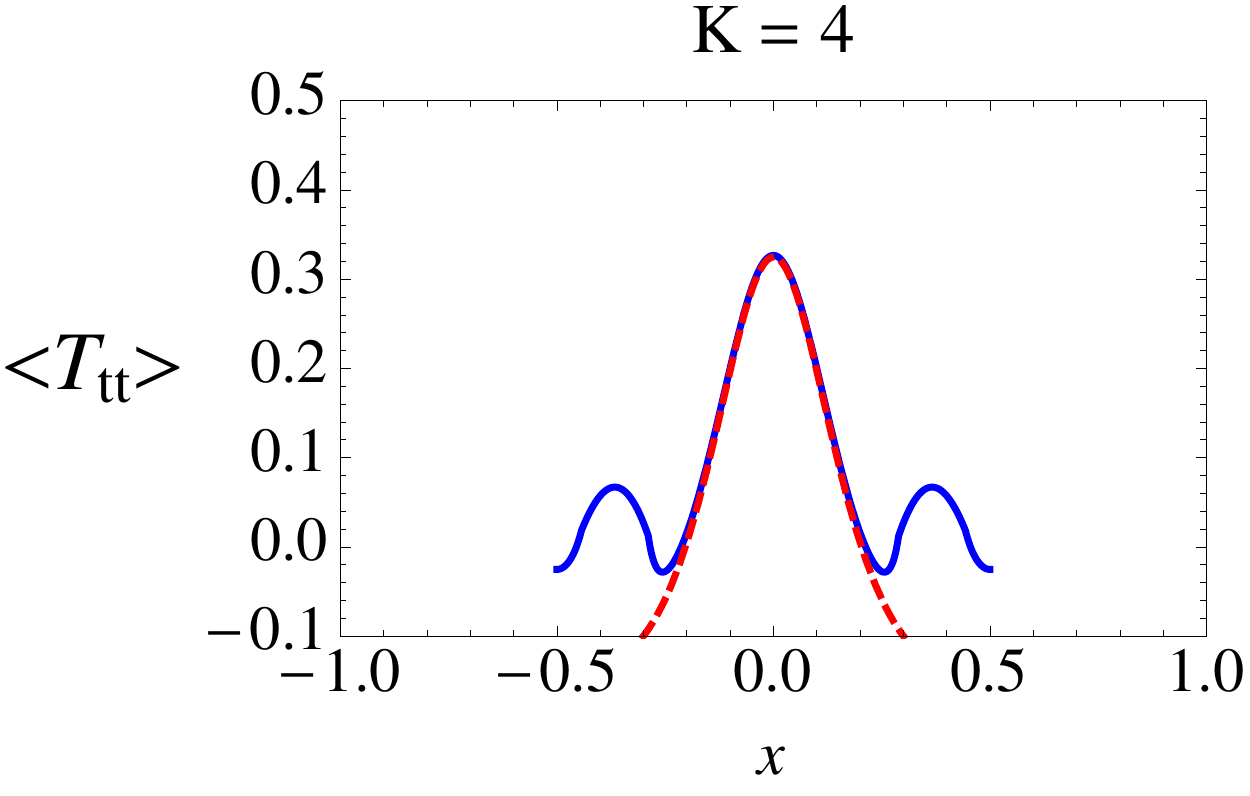}
  \includegraphics[width=0.23\textwidth]{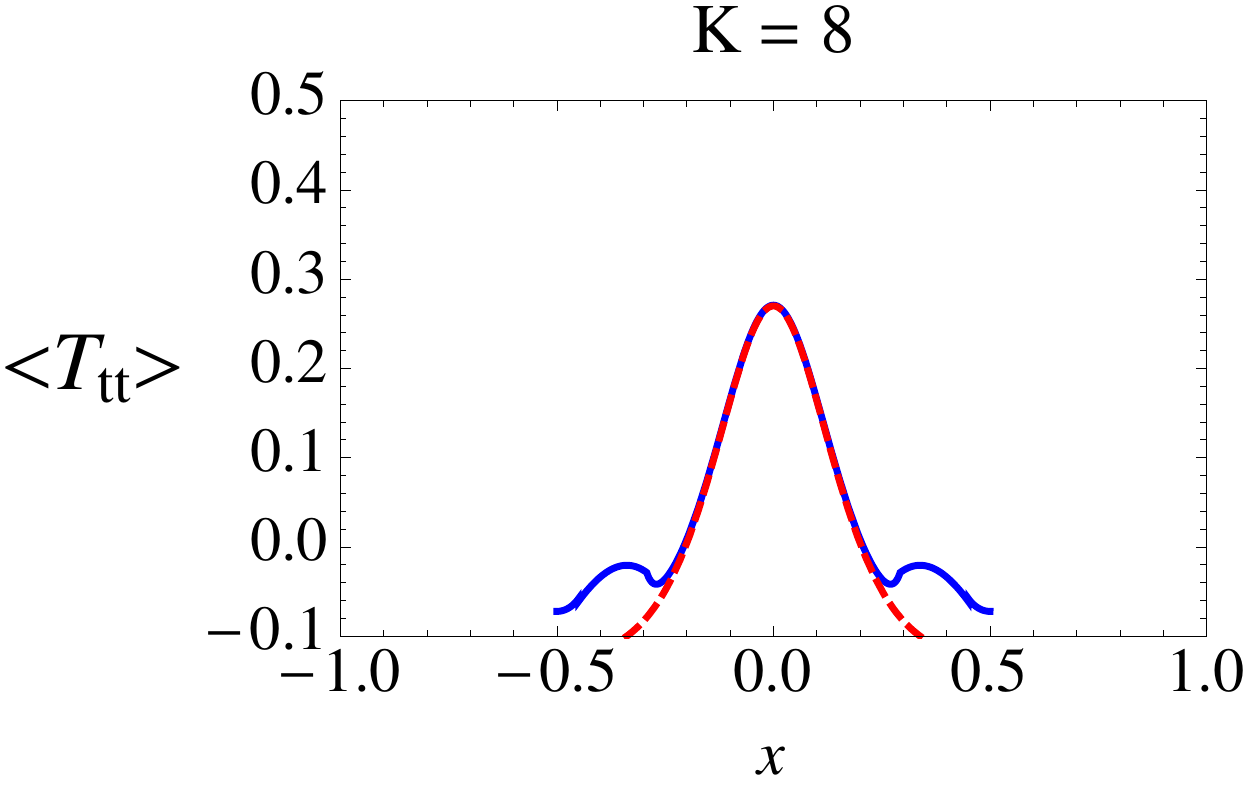}
\end{tabular}
\caption{\small{Derived regulated (see Appendix~\ref{sec:regulation}) plots for $\langle T_{tt}\rangle$ as a function of position $x$ from the analysis presented here for increasing values of Luttinger parameter $K$ at $t_x=\frac{L}{2}$ and subsequent periodic intervals ($2Lm$). Up to an overall scale factor they correspond to momentum-space distributions at $t_p=0$ and subsequent periodic intervals. Blue: full fitted distribution; dashed red: Casimir energy contribution. Top row: separation of vertex operators is $\Delta x=0.5$, bottom row: $\Delta x=0.3$. Parameters used were $L=1$ and extrapolation length $\tau_0\approx 0.278$, i.e. $k=0.9999$. The charge of the vertex operators was set to $n=1$. In the plots we have set $\phi_{0}=0$ in (\ref{eq:TtautauFull}).}}
\label{fig:mydistributions}
\end{figure*}

\begin{figure}[h!]
\vspace{0.2in}
\begin{tabular}{cc}
  \includegraphics[width=4.2cm]{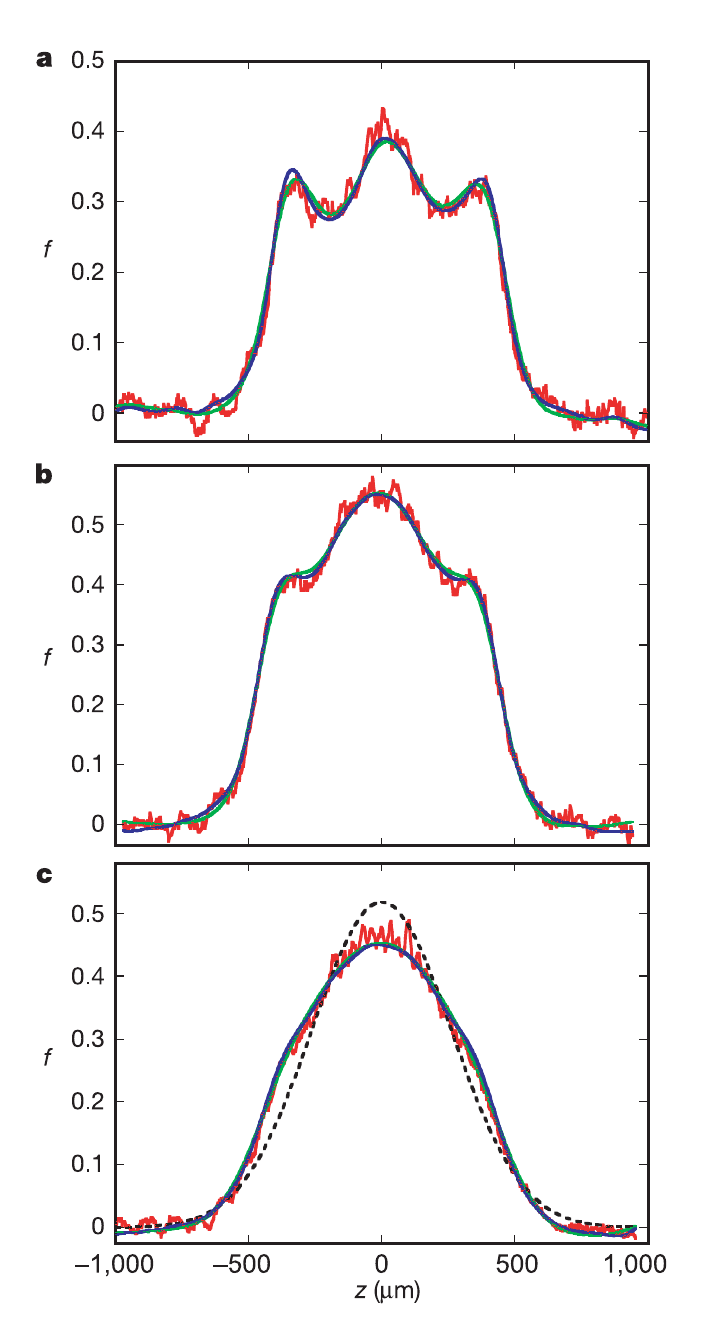}
  \includegraphics[width=4.2cm]{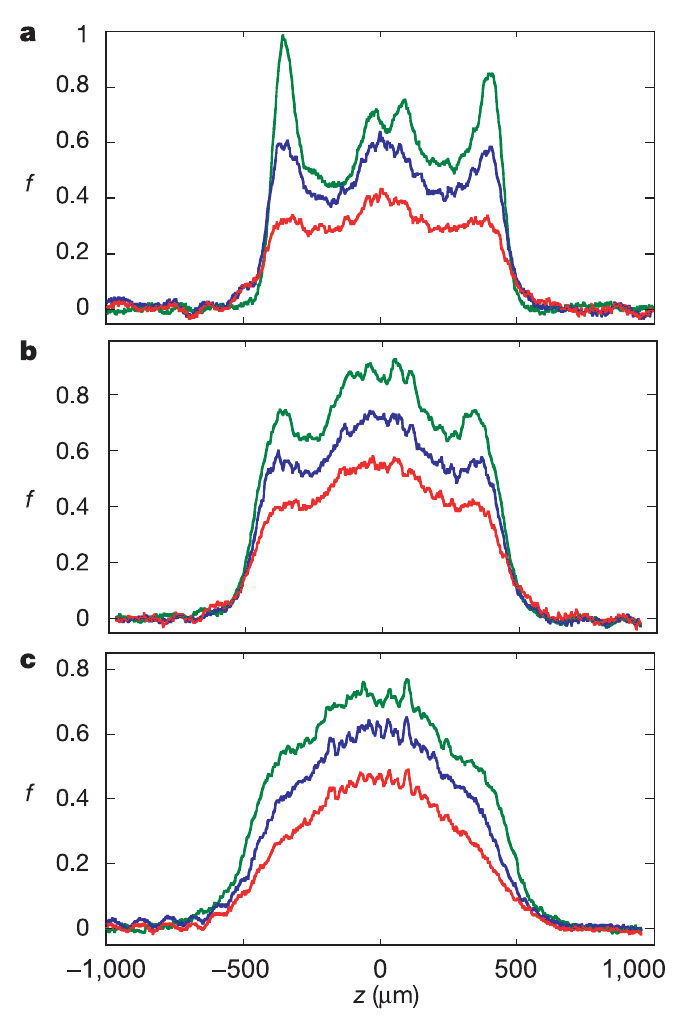}
\end{tabular}
\caption{\small{\textbf{Left:} Experimental projections~\cite{Kinoshita} of the evolution of dephased mommentum distributions without thermalization (based on observations after 15 oscillation periods) versus actual expanded momentum distributions for different dephased peaks coupling strengths $\gamma_d$ after an integer number of periods. From top to bottom: $\gamma_d=18$, $\gamma_d=3.2$, $\gamma_d=1.4$. The dashed line represents a gaussian distribution with the same number of atoms and r.m.s. width as the actual distribution. \textbf{Right:} Momentum distributions for different coupling strengths obtained by transversely integrating absorption images. Top to bottom: $\gamma=4$, $\gamma=1$, $\gamma=0.62$. Figures from~\cite{Kinoshita}.}}
  \label{fig:Kinoshitaplots}
\vspace{0.2in}
\end{figure}

Fig.~\ref{fig:mydistributions} shows the regulated $\left\langle T_{tt}\right\rangle$ plots for two such shifts corresponding to a 30\% spread and a 50\% spread respectively at $t_x$ intervals corresponding to integer-period $t_p$ intervals, and at increasing values of the Luttinger parameter. We thus see the characteristic behavior of the experimental momentum-space
distributions of~\cite{Kinoshita}, shown for comparison for decreasing
initial (input) Lieb-Liniger interaction strengths in Fig.~\ref{fig:Kinoshitaplots}. The red curves in the experimental plots are expanded momentum distributions at single periodic times and are the observable most closely expected to correspond to the derived distributions.

It would be instructive to investigate whether a more quantitative analysis could reveal deviations from the CFT description. This can be done by placing the CFT parameters in correspondence with the analogous experimental ones, such as putting the Luttinger parameter $K$ in numerical correspondence with the Lieb-Liniger interaction strengths~\cite{Cazalilla04} used in the experiment. The rectangle width, $L$, is representative of the horizontal momentum scale - but a proper fitting of the experimental plots necessitates knowledge of the original unscaled height of the experimental plots, to be compared with $\left\langle T_{tt}\right\rangle$ after rescaling by the appropriate constant to momentum space values. The extrapolation length $\tau_0$ is system-dependent, and the ability to fit this parameter could in and of itself provide interesting insight into the physics that it models. The spread of momentum in the initial peaks must also be taken into account for a proper fitting, as demonstrated in Fig.~\ref{fig:Kinoshitaplots}.

\vspace{0.1in}
\textit{Discussion ---} This intriguing qualitative agreement between the experimental distributions of~\cite{Kinoshita} and the corresponding distributions for the analogous CFT system is not expected for a general system following a quantum quench that injects high energy into the system. It may be that the experimental parameters are such that the $c=1$ CFT is still an approximate description of the system at the energies used in the experimental setup. However, the momenta injected during the quench are in principle above those that yield a post-quench Luttinger liquid. The findings of our analysis therefore call into question whether special features of the experimental system --- possibly relating to the integrability of the system --- lead to a post-quench relaxation towards a conformal fixed point.

\vspace{0.2in}
\begin{acknowledgments}
I am grateful to R. Myers for early discussions and to J. de Boer, J. Cardy, P. Chudzinski, N. Engelhardt, M. Fisher, B. Freivogel, D. Hofman, N. Iqbal, P. Kraus, M. Lippert, M. Meineri, and  A. Polkovnikov for useful comments and discussions. It is also my pleasure to thank N. J. van Druten, K. Fujiwara, E. Hudson, G. Siviloglou, and L. Torralbo Campo for their help with understanding cold atoms experimental techniques.

This work was supported by NSF grant DGE-0707424, the Netherlands Organisation for Scientific Research (NWO), and the University of Amsterdam. I am grateful for hospitality to the Visiting Graduate Fellows program at the Perimeter Institute for Theoretical Physics.
\end{acknowledgments}

\appendix
\section{Divergence regulation scheme}\label{sec:regulation}
\begin{figure}[h!]
\centering
\begin{tabular}{cc}
  \includegraphics[width=4.2cm]{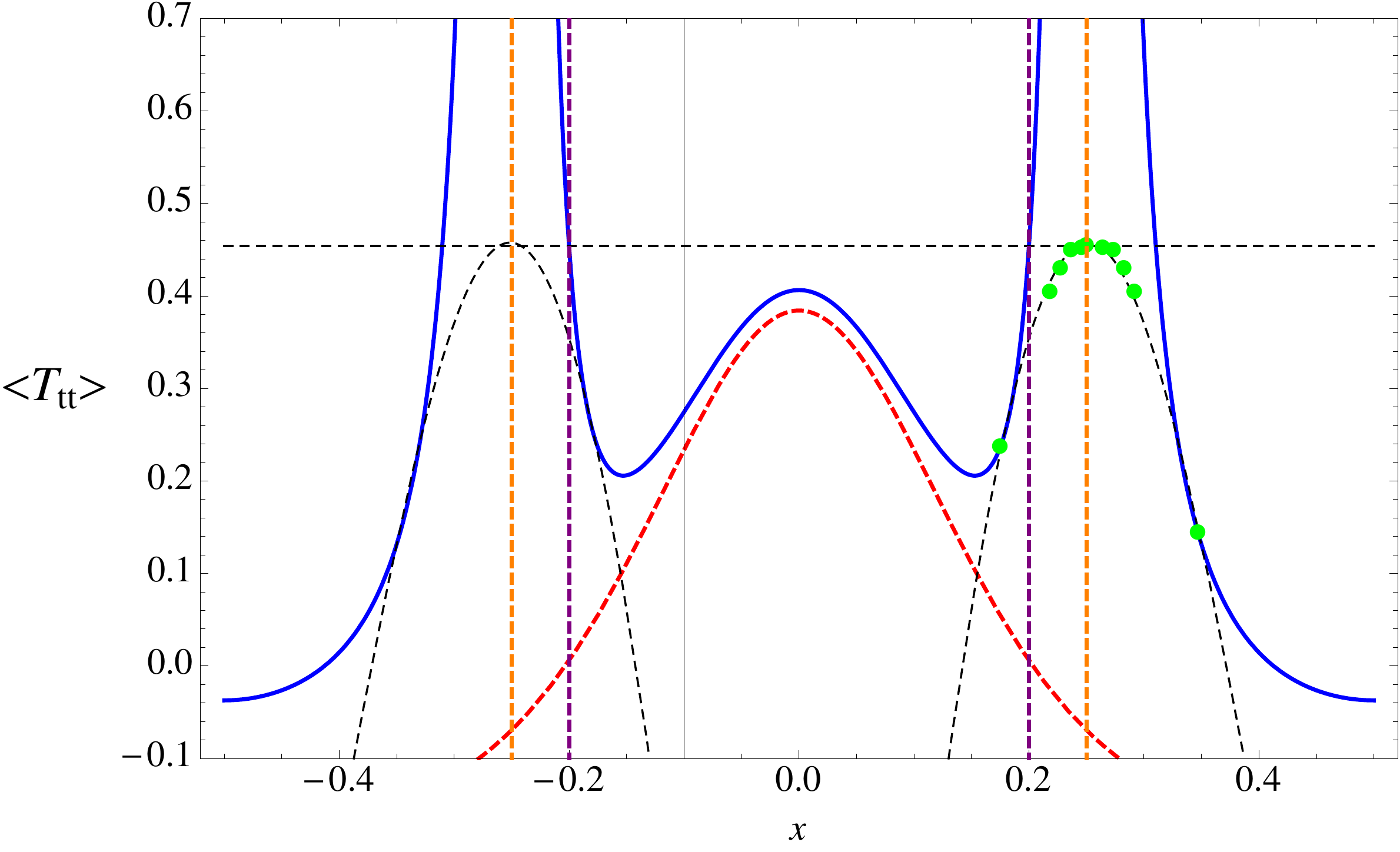}
 \includegraphics[width=4.2cm]{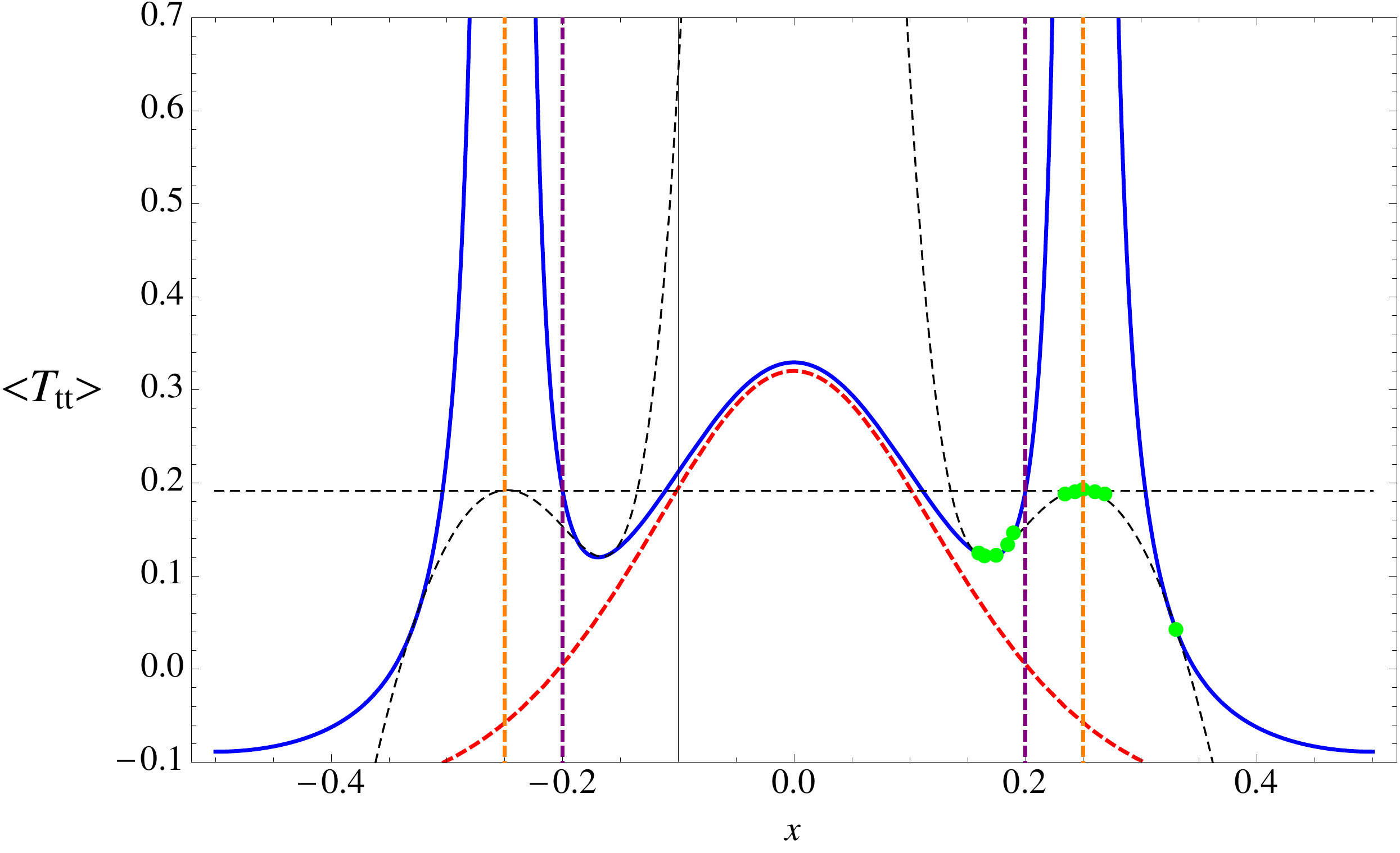}
\end{tabular}
\caption{\small{Fitting scheme for divergence regulation. Left: $K=2$, Right: $K=4$; $\Delta x=0.5$. Curves shown are unregulated $\left<T_{tt}\right>$ plot (solid blue), Casimir hump (dashed red), location of singularities (dashed orange), lower bound on critical region (dashed purple), corresponding vertical cutoff (dashed black straight line), data points used for fitting (green), and numerically fitted plots (dashed black) using a skew-normal distribution.}}
\label{fig:bestfit}
\end{figure}

In the thermodynamic limit, the correlation length $\xi\sim t^{-\nu}$, where $t=\frac{\left|T-T_{c}\right|}{T_{C}}$, diverges at the critical
temperature\footnote{We note that we use $T_c$ here for illustrative purposes; in general the particular critical parameter relevant to the system should be employed to determine the critical region of the system.} $T=T_{c}$. The size of the critical region is then given
by $t\sim\xi^{-\frac{1}{\nu}}$. In a finite-size system the correlation
length is limited by the system size; the expected scaling in a trap
of size $L$ is $\xi\sim L^{\theta}$~\cite{Campostrini}, where $\theta$
is the trap critical exponent. Experimental systems of trapped ultracold bosons in optical lattices in one dimension are well-described by the Bose-Hubbard Hamiltonian~\cite{Fisher1} and for that model it is
given by $\theta=\frac{p}{p+1/\nu}$ in the case of a power-law potential. This implies that the size of the critical region is given by $t\sim L^{-\frac{\theta}{\nu}}$.
For divergences occurring at $x=x_{0}$ we therefore place a cutoff at the height corresponding to the left boundary of the critical region, $x=x_{0}-\frac{1}{2}aL^{-\frac{\theta}{\nu}}$,
where $a$ is an arbitrary but consistent choice of constant ($a=0.1$ in Fig.~\ref{fig:bestfit}), and
round off the divergences at the corresponding height by finding a
best-fit function (skewed exponential ansatz) for a set of representative points such that a smooth choice is ensured for a given choice of $K$. We set $\nu=1$ and $p=2$ (harmonic potential) for the distributions derived here. 

\end{document}